# Efficient adaptive designs for clinical trials of interventions for COVID-19

Nigel Stallard[a,1], Lisa Hampson[a*,2], Norbert Benda[3], Werner Brannath[4], Tom Burnett[5], Tim Friede[6], Peter K. Kimani[1], Franz Koenig[7], Johannes Krisam[8], Pavel Mozgunov[5], Martin Posch[7], James Wason[9,10], Gernot Wassmer[11], John Whitehead[5], S. Faye Williamson[5], Sarah Zohar[12], Thomas Jaki[a,5,10]

[1] Statistics and Epidemiology, Division of Health Sciences, Warwick Medical School, University of Warwick, UK.

[2] Advanced Methodology and Data Science, Novartis Pharma AG, Basel, Switzerland

[3] The Federal Institute for Drugs and Medical Devices (BfArM), Bonn, Germany.

[4] Institute for Statistics, University of Bremen, Bremen, Germany

[5] Department of Mathematics and Statistics, Lancaster University, UK.

[6] Department of Medical Statistics, University Medical Center Göttingen, Germany.

[7] Section of Medical Statistics, CeMSIIS, Medical University of Vienna, Austria.

[8] Institute of Medical Biometry and Informatics, University of Heidelberg, Germany.

[9] Population Health Sciences Institute, Newcastle University, Newcastle upon Tyne, UK.

[10] MRC Biostatistics Unit, University of Cambridge, Cambridge, UK.

[11] RPACT GbR, Sereetz, Germany

[12] INSERM, Centre de Recherche des Cordeliers, Sorbonne Université, Université de Paris, France

[*] Corresponding author. lisa.hampson@novartis.com

[a] These authors contributed equally

**Abstract**

The COVID-19 pandemic has led to an unprecedented response in terms of clinical research activity. An important part of this research has been focused on randomized controlled clinical trials to evaluate potential therapies for COVID-19. The results from this research need to be obtained as rapidly as possible. This presents a number of challenges associated with considerable uncertainty over the natural history of the disease and the number and characteristics of patients affected, and the emergence of new potential therapies. These challenges make adaptive designs for clinical trials a particularly attractive option. Such designs allow a trial to be modified on the basis of interim analysis data or stopped as soon as sufficiently strong evidence has been observed to answer the research question, without compromising the trial's scientific validity or integrity. In this paper we describe some of the adaptive design approaches that are available and discuss particular issues and challenges associated with their use in the pandemic setting. Our discussion is illustrated by details of four ongoing COVID-19 trials that have used adaptive designs.

**Keywords**: Adaptive trial, Group sequential design, Multi-arm multi-stage, Platform trial, SARS-CoV-2, Pandemic research

**Word count:** 8779

**Acknowledgements:** This paper was initiated by the Adaptive Designs and Multiple Testing Procedures Joint Working Group of the Austro-Swiss (ROeS) and the German (IBS-DR) Regions of the International Biometric Society, chaired by Rene Schmidt and Lisa Hampson, and with support from Werner Brannath and Andreas Faldum. J Wason received funding from UK Medical Research Council (MC_UU_00002/6 and MR/N028171/1). This report is independent research arising in part from Prof Jaki's Senior Research Fellowship (NIHR-SRF-2015-08-001) supported by the National Institute for Health Research. The views expressed in this publication are those of the authors and not necessarily those of the NHS, the National Institute for Health Research or the Department of Health and Social Care (DHCS).

# 1. Background and the value and need for adaptive designs

## *1.1 Introduction*

Randomized controlled clinical trials are the gold standard when evaluating the potential of a novel intervention, a standard that should not be compromised during a pandemic. They are an essential component of an outbreak response (National Academies of Sciences, Engineering, and Medicine, 2017). Ideally, a rapidly conducted trial will provide definitive evidence about an intervention, allowing its immediate wide-spread deployment in the field. A vaccine that reduces risk of infection has the potential to end an outbreak. Alternatively, a treatment that improves outcomes in those who are ill may reduce transmissions by encouraging infected individuals to visit clinics (to receive treatment), thereby reducing community spread.

## *1.2 Context of trials in a pandemic*

Studies must start quickly to track with the epidemic curve and enroll enough cases. This poses a particular challenge as trials therefore need to be initiated and started before the natural history of the disease is fully understood (Baucher and Fontanarosa, 2020). In the West African Ebola outbreak, clinical trials were devised, funded and initiated in record time. However, by the time most studies opened, the incidence of new cases was dropping and most studies failed to complete enrolment (Thielman et al., 2016). A similar experience is now repeated in COVID-19 where early trials in China (e.g. Wang et al, 2020) were stopped prior to reaching the pre-planned sample sizes due to dwindling patient numbers and a large number of separate trials have been initiated investigating the same treatments. Such experience has led to some feeling that treatment trials conducted during one outbreak are about finding therapies to use during subsequent episodes of the disease, although this is certainly not always the case.

In the context of a novel infectious disease (such as Ebola or COVID-19), limited experience with treatment is common and heterogeneity of patient characteristics makes randomization an important tool for establishing intervention efficacy (Dodd et al, 2019). Additionally, careful considerations about the endpoints used in the study are required (Dodd et al, 2020). Despite these challenges, there exist recent examples of trials that immediately changed practice such as the rVSV Ebola RING vaccine trial (Henao-Restrepo et al, 2017) and PALM in Ebola (Mulangu et al, 2019).

*1.3 Aims and remit of this paper*

The recent experience in COVID-19 is that poorly conducted studies together with a certain degree of desperation mixed with endorsement of treatments by national and international leaders have meant that treatments have become widely used despite limited evidence of either benefit or safety.

This manuscript seeks to make the case that scientific rigor is equally important during a pandemic outbreak as it usually is (London and Kimmelmann, 2020). We highlight the benefits provided by the flexibility of adaptive designs. Following the recent CONSORT extension for adaptive designs (Dimairo et al, 2020; Dimairo et al, 2018), we define an adaptive design as "A clinical trial design that offers pre-planned opportunities to use accumulating trial data to modify aspects of an ongoing trial while preserving the validity and integrity of that trial". We provide a guide to the literature on statistical methods for such designs, briefly describing the main approaches and giving key references. We seek to remind readers that one must ensure that possible adaptations are considered in the final trial analysis, as some critics will otherwise raise questions about the scientific integrity of the study, and definitive trials are meant to resolve, not stir, controversy. We provide examples of COVID-19 trials that use different adaptive features and finally discuss some of the more challenging aspects of implementing adaptive designs during a major disease outbreak.

**2. Examples of COVID-19 trials with adaptive designs**

In order to illustrate some of the choices to be made when implementing an adaptive design, and the issues to be considered when making these choices, this section presents four different trials in COVID-19. These range in scope considerably, including an early (phase I/II) dose-finding platform, a confirmatory (phase III) trial, and two trials of COVID-19 patients embedded within previously-running trials. In all cases an adaptive design has been employed, the aim of which is to obtain a meaningful result from the trial as quickly as possible without any loss of scientific integrity. Our aim in including these examples is not to give a comprehensive review of COVID-19 research, or even a full description of each of these four trials, details of which can be found via the trial registration numbers given, but to highlight the nature and purpose of the adaptive design approach used in each case. More detailed descriptions of the different methodological approaches used for the adaptive designs are then given in Section 3.

*2.1 A randomized phase I/II platform trial using a Bayesian sequential phase II design: the AGILE-ACCORD platform (EudraCT 2020-001860-27)*

The first example is of a phase I/II dose-finding platform. Before investigating a new or existing compound in COVID-19 patients in a large efficacy trial, it is essential to establish safe doses and preliminary signs of activity. The AGILE-ACCORD (ACcelerating COVID-19 dRug Development, EudraCT 2020-001860-27) trial is a randomized seamless phase I/II trial platform in which multiple different candidate treatments, potentially in different populations, can be evaluated. Each candidate compound entering the platform undergoes dose-escalation initially to establish its safety profile. Safe doses are then graduated into a Bayesian adaptive group-sequential (GS) phase in which the efficacy of the doses is established. For each compound, the trial is seamless in that the efficacy data observed during the dose-escalation phase of the trial are used in the efficacy phase, and similarly safety information from the efficacy phase contributes to the safety model.

The safety of the doses is defined in terms of the risk of a dose-limiting toxicity (DLT) in the first 7 days, with dose-escalation decisions based on a randomized Bayesian model-based design (Mozgunov et al, 2019). Cohorts of patients are randomized between the control arm (Standard of Care, SoC) and the highest active doses that is safe according to the safety model. Patients within each cohort are randomized 2:1 in favor of the active dose. The inclusion of control in the dose-escalation phase is motivated by the potential overlap between DLTs and symptoms associated with COVID-19 and the need to avoid labelling potential treatments as unsafe due to misclassifying non-treatment related toxicities. Every dose that is found to be sufficiently safe can be graduated to the efficacy phase so that it is possible to have several parallel efficacy evaluations of different doses of the same compound.

The efficacy of a given dose of a compound is established via a randomized Bayesian GS trial with a time-to-event primary endpoint. The endpoint depends on the population under study. For mild and moderate cases, time to negative viral titers in nose and/or throat swab within 29 days is used while time to a 2-point improvement in the WHO clinical severity score (WHO Working Group on the Clinical Characteristics of COVID-19 infection 9-point ordinal scale) within 29 days is use for moderate and severe patients. After the efficacy data are observed, the dose can be (i) dropped for futility; (ii) graduated to a definitive trial; (iii) an additional cohort of patients recruited. Stopping is based on pre-defined stopping boundaries that have been derived under a Bayesian GS design such that the type I error for the evaluation of each dose is controlled to be below 10%. If an additional cohort is to be recruited, it will be randomized 2:1 in favor of the active dose again. The GS trial gains further efficiency by allowing up to 20 concurrent control observations from other compounds to be shared.

*2.2 A multi-arm multi stage (MAMS) trial: the RECOVERY-Respiratory Support trial (ISRCTN16912075)*

The second example is the RECOVERY-Respiratory Support trial (ISRCTN16912075). This is a multi-arm multi-stage (MAMS) randomized clinical trial to compare three non-invasive ventilation methods of standard care (oxygen given via face masks or nasal tubes), high flow nasal oxygen (HFNO), and continuous positive airway pressure (CPAP) for patients with confirmed or suspected COVID-19 who require oxygen ($FiO_2 \geq 0.4$ and $SpO_2 \leq 94\%$). The primary endpoint is the requirement of intubation or death within 30 days.

The trial team acknowledged that not all centers would be able to randomize patients between all three arms, for example due to the lack of availability of one of the interventions. To maximize recruitment to the trial it was decided to include centers provided they could randomize patients between standard care and at least one of HFNO and CPAP. Analyses will only compare patients in the HFNO (or CPAP) group with standard care patients who could have been randomized to that group, with this accounted for in the sample size.

The trial started in April 2020 with a target recruitment of 4002 patients from 40 centers. This would give 90% power to detect a reduction in the proportion of patients requiring intubation or dying from 15% to 10%. To achieve this sample size, the trial is expected to be completed within 18 months. Given the urgent need for effective treatments for COVID-19 patients and the potential long duration of the trial, the investigators wanted to stop the study early if either CPAP or HFNO was shown to be more effective than standard care, or to drop an arm or stop the study completely if one of both experimental arms was not sufficiently promising. This ability was achieved through the planning of interim analyses on a monthly basis. A formal GS stopping rule for efficacy of either experimental arm over standard care was proposed based on the alpha-spending function approach (Lan and DeMets, 1983), as described in Section 3.3 below.

The stopping rule is applied separately for each pairwise comparison with standard care. A conventional one-sided overall type I error rate of 0.025 was used, but no adjustment is planned for the multiple comparisons with standard care. No binding futility boundary was assumed, though the trial Data Monitoring Committee, who will monitor the data monthly, will be able to recommend that an experimental arm be dropped or the trial be stopped for futility at any interim analysis.

*2.3 Embedding a trial within a trial: the CAPE-Covid and the CAPE-Cod (Community-Acquired Pneumonia: Evaluation of Corticosteroids) studies (NCT02517489)*

The third example shows how an existing trial was modified to include an embedded trial of COVID-19 patients, with this embedded trial given a GS design. The CAPE-Cod trial (NCT02517489), assessing the efficacy of hydrocortisone at intensive care units (ICU) for patients suffering severe community-acquired pneumonia, was active and recruiting patients at the beginning of the COVID-19 pandemic. As SARS-CoV-2 pneumonia was not an exclusion criterion of the CAPE-Cod trial, centers started to include COVID-19 infected patients in the study. As the clinical characteristics of the two indications differ, trial stakeholders have decided to study the COVID-19 patients by embedding a specific study considering the

COVID-19 indication only, with the CAPE-Cod study interrupted temporarily to account for patients' heterogeneity in disease characteristics and evolution.

A GS design using the alpha-spending approach (Lan and DeMets, 1983; Kim and DeMets 1987a, 1987b) was used for the embedded COVID-19 study. This was chosen for its simplicity in terms of training the local methodological team, availability of software and medical community acceptance of study results. A conservative stopping boundary was selected for efficacy with an aggressive boundary for futility; that is, stopping for high evidence of superiority while stopping early if the experimental treatment is not sufficiently promising. A blinded sample size re-estimation was included in the study protocol because of uncertainty about the mortality rate for placebo patients with COVID-19, with estimates ranging from 26% to 73% (Grasseli et al, 2020; Ruan et al, 2020).

*2.4 Evaluating COVID-19 treatments in an ongoing adaptive platform trial: the REMAP-CAP trial (NCT0273570)*

The final example is of a trial that was designed to be able to adapt to an acute pandemic need. Motivated by the 2009 swine flu (H1N1) pandemic, the Randomized, Embedded, Multifactorial Adaptive Platform trial for Community-Acquired Pneumonia (REMAP-CAP, NCT0273570) seeks to: (i) evaluate multiple interventions to improve outcomes of patients admitted to an ICU with severe CAP, and (ii) provide a platform that can respond rapidly in the event of a respiratory pandemic.

The primary endpoint for COVID-19 patients entering REMAP-CAP is a composite of in-hospital death and the number of ICU free days over 21 days, thus forming a 23-point ordinal scale, i.e. -1 (death), 0, 1, …, 21. Adult patients with suspected or proven COVID-19 are enrolled into the COVID-19 stratum of REMAP-CAP, on an open-label basis, and classified into severe (in ICU) or moderate (hospitalized but not in ICU) disease states.

Within each disease state, there are multiple domains, each of which is a set of mutually exclusive interventions. An example is the antiviral domain comprising four interventions: (1) no antiviral; (2) lopinavir/ritonavir; (3) hydroxychloroquine; (4) lopinavir/ritonavir and hydroxychloroquine combination. The trial is adaptive so that new domains or interventions can be added at any time. Sites do not necessarily need to include all of the available domains or interventions.

Each patient is randomized to a regimen – one intervention from every domain – so that they receive multiple interventions simultaneously; this forms the multifactorial component of the trial design. By randomizing patients to multiple interventions, only a few receive no active treatment. Response-adaptive randomization is used to allocate patients to regimens so that more patients receive the most promising regimen(s) as the trial progresses.

Interim analyses occur frequently and are currently expected to take place weekly, although this will fluctuate with the rate of enrolment. They are also used to detect superiority, inferiority or equivalence of interventions. Depending on regional differences or treatment availability, for example, an intervention may not necessarily be dropped if it is needed in the pandemic.

**3. Statistical principles of adaptive design**

*3.1 Motivation for using adaptive design methods*

Traditionally, clinical trials testing one or more null hypotheses have followed non-adaptive designs where all details of the design are specified ahead of time. In recent decades, however, there has been a rise of adaptive designs for clinical trials in all phases of drug development

(Pallmann et al, 2018) and, in response, regulatory authorities have issued several guidelines for adaptive designs (EMA 2007; FDA 2016, 2019).

Suppose we want to perform a clinical trial to test whether an experimental treatment is superior to control (placebo or current standard of care). Measuring the benefit of the novel treatment over control by $\theta$, we want to test the one-sided null hypothesis $H_0$: $\theta \leq 0$ versus the alternative $H_1$: $\theta > 0$. In order to claim control of the type I error rate for the trial, the probability of incorrectly rejecting $H_0$ must not exceed a pre-specified level $\alpha$. The need to maintain strict control of the type I error rate is well established, particularly in the context of late-phase studies (Lewis, 1999) where a typical choice for $\alpha$ is 0.025. In many situations, however, clinical trials have several objectives. For example, trialists may want to compare several experimental treatments or active doses with control, or evaluate the effect of a novel treatment in several patient subgroups. In this setting, a trial begins testing $m$ null hypotheses $H_{0i}$: $\theta_i \leq 0$, for $i = 1, ..., m$, and often is designed to maintain strong control of the family-wise error rate (FWER) at some level $\alpha$, which is achieved if the probability of rejecting one or more true null hypotheses does not exceed $\alpha$. The requirement to control the FWER is similar to that of controlling the type I error rate in that it restricts the probability of making a false claim of efficacy.

Why should one use an adaptive design in the first place? When planning a clinical trial there might be (substantial) uncertainty about important design parameters, e.g., which effect sizes and value of the outcome variance are realistic and in which study population. If the study population is set too broadly, or the assumed variance is too small, the study will be underpowered. To address such uncertainties, it is tempting to learn from accumulating data and adapt if necessary. Adaptations of interest for COVID-19 trials may include early stopping for futility or efficacy at an interim analysis, sample size reassessment, change of allocation ratio or modification of the primary endpoint in a data dependent way.

MAMS designs (e.g., Stallard and Friede, 2008; Magirr et al, 2012; Wason et al., 2016; Ghosh et al., 2017) are of great interest for COVID-19 trials. They increase efficiency by comparing multiple experimental treatments against a shared control group (Parmar et al., 2014), while the adaptive design allows early stopping of non-promising (or highly efficacious) arms using pre-specified decision rules (Wason and Jaki, 2012). Platform trial designs (Meyer et al, 2020) allow new experimental treatments to be added as the trial progresses and often start as, or become, MAMS trials. Platform trials provide notable operational efficiency (Schiavone et al., 2019) since evaluation of a new treatment within an existing trial will typically be much quicker than setting up a new trial.

For a more detailed review of adaptive clinical trial designs, we refer the interested reader to Bauer et al. (2016) and Pallmann et al. (2018), and the references therein. For a technical description of the methods for confirmatory clinical trials, we refer to the book of Wassmer and Brannath (2016). For an overview of Bayesian adaptive methods, which have attracted much attention in the context of exploratory trials, see Berry et al (2010).

For all adaptive design approaches, data dependent changes may introduce bias and, particularly for late stage trials, appropriate statistical methods must be used to preserve the type I error rate. More details on methods to ensure error rate control is maintained are given in the next Sections, with estimation methods discussed in Section 3.5.

*3.2 Type I error rate control in adaptive designs*

All regulatory guidelines emphasize that, for confirmatory clinical trials following an adaptive design, control of the type I error rate is paramount. Elsäßer et al. (2014) reviewed scientific advice given by the European Medicines Agency (EMA) on adaptive design proposals and their outcomes (Collignon et al. 2019). The authors showed that if adaptive designs are properly implemented, then such innovative designs are well accepted by health authorities.

Strict control of the type I error rate can be achieved by using pre-specified adaptation rules such as early stopping boundaries in GS designs (Jennison and Turnbull 1999, Todd 2007) or blinded sample size reassessment rules when estimating nuisance parameters at an interim analysis (Wittes and Brittain 1990; Birkett and Day 1994; Friede and Kieser, 2006). However, pre-specified designs do not permit unplanned modifications. To allow for further flexibility, fully adaptive designs have been suggested which allow adaptations not completely specified in the adaptation rule while still controlling the type I error rate. Many designs of this type use p-value combination tests (Bauer, 1989; Bauer and Köhne, 1994) to allow for more flexible decision making, while still preserving the integrity of the design. Alternatively, designs follow the conditional error principle (Proschan and Hunsberger 1995; Müller and Schäfer 2001, 2004) in order to control the type I error rate. The conditional error principle enables one to introduce flexibility to any type of pre-defined design, even if no interim analysis was originally planned (Müller and Schäfer, 2004); the key is that the significance level of the adapted trial must not exceed the level of the initial design conditional on the data already observed. If data are combined across stages of the adaptive trial using the inverse normal function proposed by Lehmacher and Wassmer (1999) and Cui et al. (1999), and monitored against GS boundaries, the test decision should coincide with the study outcome had no unplanned adaptation been performed.

If more than one null hypothesis is to be tested, for example in a multi-arm study, several extensions (Bauer and Kieser, 1999; Hommel, 2001) to the designs discussed above have been proposed which use adaptive designs within a closed testing framework (Marcus et al. 1976). A closed testing procedure requires level α tests of each individual null hypothesis to be specified, as well as level α tests of all possible intersections of null hypotheses; if all tests involving a specific hypothesis are significant at level α, then that null hypothesis may be rejected globally at level α. Tests maintaining strong control of the FWER must (whether explicitly defined or not) be a closed testing procedure (Burnett, 2017). For adaptive trials, for each null hypothesis, whether individual or an intersection, an adaptive test is pre-specified.

The advantage of the fully adaptive design is that both internal and external information can be used to adapt the trial without fully specifying the adaptation rule. This is crucial in the pandemic situation as it means emerging data from parallel trials can be easily incorporated into the adaptation decision rule of an ongoing adaptive trial without compromising its integrity.

The methods discussed above focus on ensuring type I error rate control in a frequentist setting. It is also possible to incorporate Bayesian methodology into the decision making process for adaptations without compromising error rate control (Berry et al, 2011; Stallard et al., 2020).

### *3.3 GS stopping rules in confirmatory adaptive design trials*
Incorporating a GS stopping rule into a confirmatory clinical trial allows for greater flexibility as compared to a standard trial design without any interim data looks. If an extraordinarily large benefit of the experimental treatment over control is observed at an interim analysis, it is desirable to have the option to stop the trial for efficacy with early rejection of the null hypothesis. In case such an option is incorporated, a multiple testing problem emerges which needs to be handled by adjusting the local significance levels for the interim and final analysis in order to maintain the nominal type I error rate. In GS trials with one or more interim looks, this can be done by using adjusted rejection regions tailored to pre-specified interim analysis times (O'Brien & Fleming 1979; Whitehead and Stratton, 1983). In practice, many GS trials are run using the alpha-spending function approach (Lan & DeMets 1983; Kim & DeMets 1987a) or other methods (Whitehead, 1997a) that offer more flexibility with regards to the timing and number of interim analyses (as long as these are chosen based on blinded data). GS stopping rules for efficacy can also be incorporated in adaptive designs using a combination test approach (Bauer & Köhne 1994, Lehmacher & Wassmer 1999) or the conditional error principle (Proschan & Hunsberger 1995, Müller & Schäfer 2001). Alternatively, one might also consider stopping a confirmatory trial early for futility if data are consistent with clinically irrelevant or negative treatment effects, in order to prevent harm to patients and to enable resources to be diverted to more promising interventions. A futility rule can either be binding in the sense that it needs to be strictly adhered to at the interim analysis, or non-binding, meaning that the rule may also not be followed. While binding futility rules allow for a relaxation of the local significance levels, non-binding rules can be regarded as more advantageous from a practical perspective due to their increased flexibility. However, the type II error rate increases with the use of a non-binding futility rule, requiring an increased sample size to achieve the target power.

### *3.4 Multi-arm multi-stage (MAMS) and platform trials*

MAMS designs allow several experimental treatments to be assessed in a single trial with treatments dropped from the trial as soon as interim analysis data suggest they are ineffective. The design requires specification of stopping boundaries prior to the trial starting: these usually consist of futility stopping boundaries and, less commonly, efficacy stopping boundaries. At an interim analysis, each experimental arm is compared against control using all patients with outcome information up to that point: experimental arms with test statistics below the futility boundary have further recruitment stopped and experimental arms with test statistics above the efficacy boundary can be concluded to be effective (and the respective null hypothesis rejected). Setting stopping boundaries is not trivial, but can be done using analytical formulae (Royston et al., 2003; Magirr et al, 2012) or simulation (Wason and Jaki, 2012). Software exists such as the R package MAMS (Jaki et al., 2019) and the Stata module nstage (Bratton et al., 2019). The stopping and selection rules are pre-planned, meaning that the trial must be conducted according to the pre-specified rules. This can be generalized to a flexible confirmatory adaptive approach (Magirr et al., 2014) which is usually based on the combination testing principle (Bauer, 1989). This principle can be applied for testing multiple hypotheses by use of the closed testing principle (Bauer et al., 2016; Bauer and Köhne, 1994; Bretz et al., 2009; Wassmer and Brannath, 2016; Posch et al., 2005). An application is the multi-arm setting where, for example, treatment arms can be selected and sample sizes for the selected treatment arms can be recalculated based on the observed response (Bauer and Kieser 1999; Posch et al. 2005; Koenig et al 2008; Magirr et al 2014; Wassmer and Brannath, 2016). For these designs, the adaptation rules, need not be pre-specified. Consequently, this approach allows for testing in MAMS trials controlling the FWER in a strong sense. R packages are available for confirmatory adaptive designs in the multi-arm setting, such as asd (Parsons et al., 2012; Friede et al, 2020) and rpact (rpact, 2020), with MAMS (Jaki et al., 2019) also having an extension for dealing with unexpected design modifications.

Some MAMS trials use adaptive randomization (AR) to guide allocation to better performing arms. In a multi-arm setting, if the shared control group is kept separate from the AR procedure, this maintains the power at a high level (Trippa et al., 2012; Wason and Trippa, 2014; Williamson and Villar, 2020; Viele et al., 2020). Frequentist and Bayesian approaches to AR can be used, although in practice it is more common to use a Bayesian approach. AR has been criticized for being susceptible to temporal trends in the trial (Thall et al. 2015, Proschan and Evans, 2020), which may be possible in COVID-19 settings due to mutation of the virus. However, if the allocation to the shared control group is maintained, there is little impact on the statistical properties of the trial (Villar et al., 2018).

Another type of MAMS design is the drop-the-losers design (Sampson and Sill, 2005; Wason et al, 2017), which sets a fixed number of arms to progress at each interim analysis. An advantage of this is that the design has a fixed sample size (in contrast to a random sample size for the approach described above). The main disadvantage is that it may lead to dropping of promising arms if there are more than the design is permitted to take forward. Approaches that combine the drop-the-losers approach with pre-specified stopping boundaries have been proposed (Friede and Stallard, 2008) to allow early stopping for futility and efficacy.

Magirr et al. (2012), Wason et al (2017) and Robertson and Wason, (2019) describe how FWER control can be achieve in a MAMS design. However, there has been debate about which type of multiplicity adjustment, e.g. pair-wise or family-wise, is needed for a multi-armed design (Wason et al., 2014; Howard et al. 2018; Stallard et al. 2019; Collignon et al. 2020). Generally, the decision will be determined by the relative importance of making a type I error and a type II error, as correcting for multiple testing reduces the power. In a platform setting, it is more difficult to formally control the FWER unless it is known in advance how many treatments will be tested. New methods for online control of error rates may provide a way of doing this if it is required (Robertson and Wason, 2018).

Most methodology work in MAMS has focused on settings with a shared control group. In the case of no obvious control group, we would refer the reader to the Magaret et al. (2016) and the motivating example in Whitehead et al. (2020).

When there is a shared control group, the relative allocation between control and experimental treatments can affect the power. A natural choice is to randomize patients equally between each available treatment. One can gain a small amount of power by increasing the allocation to the control group (Wason et al. 2016; Wassmer, 2011; Viele et al, 2020), although in a MAMS study the optimal allocation depends on the likelihood of dropping arms only. Setting the allocation ratio of new arms in a platform study can also be optimized (Bennett and Mander, 2020).

*3.5 Estimation after an adaptive design trial*

Point estimates and confidence intervals (CIs) for the treatment effects assessed in a confirmatory clinical trial play a key role in deciding whether or not to recommend adoption of the new interventions in clinical practice. Therefore, the estimators used in confirmatory adaptive clinical trials need to have good accuracy. Naive point estimators that assume a two-arm single-stage design can be substantially biased for an adaptive trial because of the possibility of stopping early for futility or overwhelming effect (Whitehead, 1986; Koopmeiners et al., 2012) and/or selecting the most promising experimental treatments at an interim analysis (Bauer et al, 2010; Carreras and Brannath, 2013; Kimani et al., 2013). For two-stage trials, conditional on the trial continuing to stage 2, the selection rule and the treatments continuing to stage 2, several uniformly minimum variance conditional unbiased estimators

(UMVCUEs) exist (Kimani et al., 2013; Cohen and Sackrowitz, 1989; Bowden and Glimm, 2008; Robertson et al., 2016; Stallard and Kimani, 2018). The UMVCUEs can have large mean squared errors (MSEs) and are computed if the trial continues to stage 2. Alternative shrinkage estimators that are based on empirical Bayes approaches tend to have smaller MSEs but they are not necessarily unbiased (Carreras and Brannath, 2013; Brueckner et al., 2017). Estimators that also reduce bias when a trial stops in stage 1 exist (Carreras and Brannath, 2013; Shen, 2001; Stallard and Todd, 2005) as well as generalizations to multiple stages (Whitehead et al., 2020).

Naive CIs can have incorrect coverage. When an adaptive design selects the most effective experimental treatment in stage 1, among the methods that adjust for adaptation (Stallard and Todd, 2005; Sampson and Sill, 2005; Wu, Wang and Yang, 2010; Magirr et al., 2013), the Sampson and Sill (2005) method performs best in terms of coverage while the approach of Wu et al. (2010) is the most straightforward for computing CIs (Kimani et al., 2014). Methods for computing CIs with other selection rules are available (Magirr et al., 2013, Neal et al., 2011). One of these methods allows continuing with multiple experimental treatments in stage 2 but sometimes the computed CIs are non-informative (Magirr et al., 2013).

Most of the methods above assume normally distributed outcomes with known variance. Extensions to the case of unknown variance (Robertson and Glimm, 2019) exist and, using asymptotic approximations, they can be extended to other outcomes. Point estimation with time-to-event outcomes has been considered separately as these outcomes pose specific challenges (Brückner, Titman and Jaki, 2017). One of the UMVCUEs is a generalization of the others enabling MAMS trials (Stallard and Kimani, 2018).

## 4. Specific statistical issues in adaptive designs in COVID-19 intervention trials

### *4.1 Characterizing the dose-response relationship*

When studying a compound in a completely new disease population, such as COVID-19, it is essential to identify a therapeutic window, that is, a range of doses which simultaneously control the risk of adverse events while having a clinically meaningful efficacy effect. Importantly, the same requirement applies to repurposed medicines as the dose selected for the original indication is not guaranteed to be optimal for the COVID-19 population. Characterization of the dose-response relationship, however, poses several statistical challenges.

Both the toxicity and efficacy signals should be attributable to the compound rather than to the baseline toxicity and efficacy rates. Therefore, the dose-response relationship should be established accounting for outcomes on control, by quantifying differences between the active doses and control (standard of care or placebo) for both toxicity and efficacy endpoints. Furthermore, to gain efficiency by using data across different doses of the compound, a model-based approach that stipulates a parametric relationship for the underlying differences should be used. This, in turn, poses a challenge in terms of how to select the dose-response model. To overcome this, an MCP-Mod approach which considers several candidate models (Bretz et al, 2005; Pinheiro et al., 2014), or a model-averaging approach (Buckland et al, 1997) can be used. To establish the dose-response relationship, randomization of patients between doses is needed to allow for efficient dose-response estimation. Consequently, the safety of the dosing range must be established before the dose-response relationship can be reliable characterized. A two-stage procedure adjusting allocation probabilities to each dose could also be used (Mielke and Dragalin, 2017).

To ensure the most efficient use of the data, the most informative yet clinically meaningful endpoint should be used. While it is common to dichotomize continuous or time-to-recovery endpoints to simplify the analysis, it is inevitable that this will result in the loss of some statistical information; specifically, this strategy has been found to result in noticeable losses when characterizing the dose-response relationship (Mozgunov et al., 2020). While dose-response models for continuous and binary endpoints are more common, there are also parametric and semi-parametric models available for time-to-event data with implementations available in standard statistical software; see, for example, Laga and Boateng (2020) for an implementation of MCP-Mod for survival data.

*4.2 Stopping a trial early due to recruitment difficulties*

The incidence of a disease will vary over the course of a pandemic. A trial starting recruitment during the 'post-peak' phase of a pandemic, as disease activity levels trail off, may struggle to reach the target information level before the pandemic ends. If a GS test is stopped early due to feasibility issues, steps can be taken to draw valid inferences about the target treatment effect. As mentioned in Section 3.3, many GS trials are monitored according to the alpha-spending function approach (Lan and DeMets, 1983), and accommodating early termination in this framework is relatively straightforward. At the final analysis, we choose the critical value of the test so as to 'spend' all the remaining type I error rate. Choosing the final critical value in this way preserves the type I error rate of the trial, although there will be a drop in power relative to the originally planned design. CIs and p-values based on the stagewise ordering of the test's outcome space can be calculated conditioning on the observed sequence of information levels (Fairbanks and Madsen, 1982, Tsiatis et al., 1984). A similar idea can be used to reach a final hypothesis decision in a GS trial which does not follow an alpha-spending design: one finds the critical value for the final analysis so that the trial has type I error rate $\alpha$ given the observed information levels and the test boundaries used at previous interim analyses. One additional caveat that applies to all GS trials stopped early for lack of feasibility: while the decision of when to truncate the GS test can (and should) be informed by trends in recruitment rates, the timing of the final analysis should not be influenced by previous treatment effect estimates (Jennison and Turnbull, 1999). If, for example, we bring forward the final analysis because we were close to crossing a boundary at the last interim look, test statistics will not follow the anticipated canonical joint distribution used to calculate stopping boundaries (Jennison and Turnbull, 1997; Scharfstein et al., 1997), leading to deviations from the nominal type I error rate.

If the pandemic ends earlier in some geographic regions than others, the sponsor of a global study can compensate by switching to recruit more patients at sites where the disease is still active. While this strategy would preserve the type II error rate of the trial, one would still see perturbations in the power of region-specific subgroup analyses. In addition, the impact of this shift in recruitment on the interpretation of the trial should be carefully considered, particularly if the target estimand remains as the effect of treatment in a global patient population (which would still be pertinent if all regions remained at risk of subsequent waves of infection). To align with this target estimand, the analyst could perform a random effects meta-analysis of region-specific treatment effect estimates, where the mean of the random-effects distribution would be taken as the overall treatment effect of interest.

### *4.3 Heterogeneity of the patient population*
The prognosis of COVID-19 is highly variable, ranging from non-symptomatic courses of the disease in the majority of patients, to severe clinical courses requiring hospitalization, intensive care and leading to death in many cases. A number of prognostic factors have been identified, including age, sex, chronic lung disease, diabetes, hypertension and other cardiovascular comorbidities (Zhou et al., 2020; Chen et al., 2020; Grasselli et al., 2020; Jordan et al., 2020). In addition, the time span and course of the disease up to the time of study inclusion is prognostic for the outcome. For the design of clinical trials this implies that the inclusion criteria have a strong impact on the distribution of the primary endpoint and the corresponding sample size planning. Furthermore, in the analysis of the clinical trial, the precision of treatment effect estimates can be substantially improved by adjusting for important covariates in the analysis. Also, stratified randomization according to the risk factors can improve the efficiency of the design.

An important question is to what extent the prognostic factors are also predictive for the efficacy of the treatment. For example, for anti-viral drugs it is usually assumed that efficacy is largest when treatment starts early after infection. However, co-morbidities and demographic variables can also play a role in the mechanism of action or affect the safety profile. If the treatment effect varies across subgroups that are defined by baseline characteristics, treatment effect estimates are only meaningful for a well-defined patient population. In addition, subgroup analyses will be critical to better understand the treatment effect. Challenges are the typically small sample sizes in subgroups and the multiplicity problem connected with multiple analyses. Adaptive enrichment designs can be a useful tool in this context (see e.g. Friede et al, 2020 for a recent overview). Based on interim results, they allow one to increase the sample size in specific subgroups or to restrict randomization to these populations, while controlling frequentist error rates (Ondra et al. 2019, 2016; Chiu et al., 2018). However, pre-specification of subgroups can be difficult, especially for quantitative predictive factors (e.g. time from the onset of the disease), where thresholds need to be pre-specified in order to define the corresponding sub-populations. If there is not enough prior information to select a single threshold, subgroup analyses with multiple thresholds can be implemented if an appropriate adjustment for multiple testing is foreseen (Graf et al., 2019, 2020).

*4.4 Dynamically changing standard of care*

SoC is likely to evolve quickly in the early stages of a pandemic, particularly of a previously unknown disease, as clinicians' understanding of the disease and experience treating it increases. Of course, many trials in an acute pandemic setting recruit quickly and measure short-term endpoints, such as survival at 28 days, meaning that SoC would be expected to remain relatively stable over the course of an individual study. However, more substantial changes may be seen over the course of a trial measuring a long-term outcome. Alternatively, we may see changes in SoC across the course of several successive studies which we intend to combine via meta-analysis.

For trials of add-on treatments (comparing SoC + novel therapy versus SoC alone), block-randomization or pre-specification of the allocation probabilities in a parallel group study will be sufficient to ensure control of the type I error rate at the nominal level. If changes in SoC result in a time-varying treatment effect, the study investigators should consider what the target estimand is, that is, an aggregate treatment effect averaging across the course of the study, or the effect of treatment compared with the most recent version of SoC, and whether the target effect can be estimated.

Changes in SoC can present a particular challenge in a platform trial adding in new experimental treatments over time. In this setting, there is debate about whether one should use non-concurrent control patients in the evaluation of experimental treatments. If a new treatment is added in part-way through the trial, concurrent control patients are those who are recruited from that time; non-concurrent controls would be those patients recruited prior to the addition. If it is likely that there is strong potential for a temporal trend in the data, as may be the case in COVID-19 trials, it is likely best to use only concurrent controls as the primary analysis.

*4.5 Leveraging information across trials using meta-analytic approaches*

Comparison-specific treatment effects may accumulate over the course of a pandemic as randomized controlled trials are replicated by different research groups, and as platform trials with common treatment arms report pairwise comparisons. Combining data across studies in a meta-analysis will increase the statistical information available for estimating a target treatment effect. If we prospectively intend to update the meta-analysis as each new study is published, the cumulative results should be monitored using a GS test in order to avoid over-interpretation of random highs and lows in treatment effect estimates, and inflation of the type I and type II error rates (Berkey et al., 1996). Different approaches have been proposed for monitoring a cumulative meta-analysis (Simmonds et al., 2017). Sequential meta-analysis approaches use GS boundaries to monitor standardized Z statistics obtained from random-effects meta-analyses. For example, boundaries could be from a triangular GS test or a restricted procedure using a 'Christmas tree' correction for discrete monitoring (Whitehead, 1997b, Higgins and Whitehead, 2011), or an alpha-spending test spending type I error as a function of the estimated proportion of the target information level accumulated thus far. The DerSimonian-Laird (1986) estimate of the between-study heterogeneity may be unstable in the early stages of a cumulative meta-analysis when few studies are available, so that estimated information levels may decrease between successive interim analyses (Higgins and Whitehead, 2011). With (very) few studies also Bayesian approaches with weakly informative priors for the between-study heterogeneity were found to be more robust than standard methods (Friede et al, 2017). It may be possible to reduce between-study heterogeneity and increase borrowing across studies by leveraging data on baseline covariates, either in a meta-regression approach if only aggregate data are available, or in models for patient responses if individual patient data are available.

**5. Practical issues in conducting adaptive designs**

As highlighted in Section 1, time is essential when conducting clinical trials during a disease outbreak. When conducting an adaptive trial, one particular aspect, namely decision making about undertaking any of the (pre-planned) adaptations, needs to be particularly considered in addition. This time-sensitivity means that the decision processes (which often involve review of unblinded data) are streamlined and decision making is clear before undertaking any interim data looks to avoid potential delays in implementing decisions. The data monitoring committee (DMC) of the trial will often have a particularly crucial role to play in these decision processes.

*5.1 Data monitoring committees for an adaptive trial*

A DMC, also known as data safety monitoring board (DSMB) or data safety monitoring committee (DSMC), is "a committee created by a sponsor to provide independent review of accumulating safety and efficacy data" (Herson, 2009). Therefore, the DMC is "responsible for […] protecting trial integrity and patient safety" (Herson, 2009). Although all trials require safety monitoring, this does not mean that a DMC needs to be established for all trials. However, safety monitoring by a DMC is recommended "for any controlled trial of any size that will compare rates of mortality or major morbidity" or more specifically for a "population at elevated risk of death or other serious outcomes, even when the study objective addresses a lesser endpoint" (FDA, 2006), which is likely to be the case for any COVID-19 trial. Compared to other studies, COVID-19 trials might be shorter with faster recruitment (given the pandemic situation) and shorter length of follow-up (e.g. restricted to length of stay in hospital). For safety monitoring by a DMC this, in turn, requires shorter review intervals, which might lead to weekly safety reviews in some instances. This has obvious consequences for logistics in terms of data management, statistical analyses, as well as the required availability and commitment of the DMC members. Given these demands it might be practical to set up committees to monitor not only an individual trial but to oversee several trials on a programme level. This would not be uncommon for trials investigating the same compound, but in COVID-19 might be applied also to settings where different compounds are being investigated by the same sponsor. This would also enable the DMC to consider data emerging from other ongoing trials in their decision making, allowing them to integrate rapidly increasing and changing information. This type of oversight committee tends to be larger than the standard DMC which typically comprises two physicians and one statistician, since a wider range of expertise might be required and several regions might need to be represented. In the case of adaptive designs or complex platform trials, the DMC might be charged with extra responsibilities beyond safety monitoring. For example, when performing a formal interim analyses, the DMC might make recommendations regarding treatment or subgroup selection, as well as futility stopping or sample size re-estimation. Since COVID-19 trials often use quickly observed endpoints, such adaptations might still be meaningful, although the trials might be quite short overall given rapid recruitment.

*5.2 Regulatory considerations*

Basic regulatory requirements on adaptive designs to be used in confirmatory study are laid down by the FDA and EMA (EMA, 2007, FDA 2016, 2019). In general, regulatory concerns in general relate to the validity and robustness of the conclusions to be made with respect to a statement on the drug's efficacy and a thorough benefit risk assessment. In statistical terms, proper type I error control and unbiased effect estimates (or at least minimization of bias) are of major concern for the primary efficacy endpoint(s), whereas the potential lack of mature data in trials with interim decision relate to the totality of the data package. In addition to proper statistical methods to control type I error and bias in estimation, issues with the integrity of the trial due to change in trial conduct following an interim evaluation (and leading to inconsistency between stages) play a major role in the assessment of the trial's validity.

Referring to the specific pandemic setting, the request for rapid decisions, investigation of multiple treatments, and potentially changing controls may complicate the multiplicity issues and the assessment of the trial's validity on one hand. One the other, the high medical need for effective drugs may also lead to a different appraisal of type I error rate control to make new treatments available as fast as possible. Nevertheless, it remains paramount to understand the trial's operating characteristics to ensure a fully informed regulatory decision.

In the specific setting of the COVID-19 pandemic with a number of highly relevant risk factors, high heterogeneity with respect to the severity of the disease and the onset of treatment, a good characterization of the population that would profit from treatment is crucial. Obviously, an extensive search for the optimal subgroup and subsequent confirmation may not be compatible with the need for rapid development. Regulatory principles of subgroup selection are given in the EMA guideline on the investigation of subgroups in confirmatory clinical trials (EMA, 2019), to be applied in case a positive effect has been established in the overall population but plausible heterogeneity with respect to efficacy and severe side effects asks for a refinement of the population.

## 6. Conclusions and discussion

Although there can be no doubt that clinical research in the setting of a pandemic raises a number of unique challenges, the need for the evaluation of potential treatments in scientifically rigorous randomized controlled clinical trials is as great as in any other area of clinical research.

Whilst it is always desirable to conduct clinical trials of new medical interventions in a timely fashion, faced with a pandemic for which no vaccines or treatments are available, this need is particularly acute, both because of the urgent need for treatments and because of the limited time-window in which large-scale research may be possible in a particular location, or even globally.

Although trials that are unable to fully enroll in time to inform a current outbreak may provide crucial data for future outbreaks or may be analyzed in the context of a meta-analysis, the need for clinical trials that can yield a conclusion, either positive or negative, as quickly as possible, makes adaptive design methods particularly attractive, especially if they are flexible enough to allow modifications to a trial in light of emerging clinical knowledge and results from other ongoing or completed clinical trials. It should be possible to suspend recruitment to a trial during times when the disease is under control and to resume if it re-emerges (Dean et al., 2020). The use of adaptive platform designs, in particular for clinical trials in COVID-19, has been proposed by a number of authors (Baucher and Fontanarosa, 2020, Murthy et al., 2020) and, as illustrated by the examples given in Section 2, a range of sequential and adaptive design methods are being employed in ongoing clinical trials. Although not a focus of this paper, adaptive design methods also provide valuable tools to allow for modification of trials in other diseases necessitated by the COVID-19 pandemic (Kunz et al, 2020).

In order to yield timely results, it is clear that treatment trials should start as soon as possible after an epidemic has begun. The time limiting factor should be identification of candidate treatments, and not the development of a statistical design. Hence, research on trial designs that are most suitable for the purposes of a pandemic should become a research topic that goes beyond the actual case in order to be better prepared for future pandemics.

A suitable design, or small suite of designs, should be available on the shelf ready for use at any time. Debates about trial design, including whether "best available care" can vary amongst centers and whether Bayesian or frequentist approaches are appropriate should be settled between pandemics, and not re-opened when they strike. It might be supposed that such a generic approach is impossible and that the design cannot be created without details of the natural history of the disease or the specification of candidate treatments. However, in almost all cases the primary patient response will be status, a certain number of D days after randomization, classified as dead/severely ill/moderately ill/recovered. The magnitude of differences between treatments can be conveniently assessed in terms of odds-ratios, and an odds-ratio of 2 is almost always clinically important. A potential generic design has been suggested by Whitehead and Horby (2017, 2018) that might be a starting point in the search for a suitable vehicle. Perhaps the COVID-19 experience will motivate the clinical trials community to devise and agree upon such a generic design once the current pandemic subsides. It should accommodate the evaluation of multiple treatments compared with best available care, with treatments entering the study as they are devised and are approved for experimental use, and leaving as they are demonstrated to be inferior to others. The design should be adaptive and capable of being used across many sites, preferably internationally.

Even if suitable trial designs are available, the planning of clinical trials remains challenging when not only healthcare resources are overstretched but the number of patients likely to be affected as well as the duration of the pandemic is unknown and may vary considerably across the globe. For rigorous and quick progress in evaluating treatments, there is a need for scientific collaboration, and the necessary political collaboration to enable and facilitate this. A multi-center trial can recruit patients as quickly as possible by recruiting internationally, or from different countries at different times as the pandemic progresses. In a setting where scientific knowledge is rapidly changing and research studies are being set up quickly, data sharing among global scientific communities is key in allowing meta-analysis methods, such as those described in Section 4.5, to be used. Consequently, results from a number of studies can be combined to provide robust evidence of treatment effectiveness, thus preventing wastage of resources arising from the conduct of multiple similar trials, each of which are individually too small to yield definitive results.